# Silicon-Organic Hybrid (SOH) Mach-Zehnder Modulators for 100 GBd PAM4 Signaling With Sub-1 dB Phase-Shifter Loss

CLEMENS KIENINGER,[1] CHRISTOPH FÜLLNER,[1] HEINER ZWICKEL,[1] YASAR KUTUVANTAVIDA,[1] JUNED N. KEMAL,[1] CARSTEN ESCHENBAUM,[1] DELWIN L. ELDER,[2] LARRY R. DALTON,[2] WOLFGANG FREUDE,[1] SEBASTIAN RANDEL,[1] AND CHRISTIAN KOOS,[1,3*]

[1]*Karlsruhe Institute of Technology (KIT), Institute of Photonics and Quantum Electronics (IPQ), 76131 Karlsruhe, Germany*
[2]*University of Washington, Department of Chemistry, Seattle, WA 98195, USA*
[3]*Karlsruhe Institute of Technology (KIT), Institute of Microstructure Technology (IMT), 76344 Eggenstein-Leopoldshafen, Germany*
*christian.koos@kit.edu*

**Abstract:** We report on compact and efficient silicon-organic hybrid (SOH) Mach-Zehnder modulators (MZM) with low phase shifter insertion loss of 0.7 dB. The 280 µm-long phase shifters feature a π-voltage-length product of 0.41 Vmm and a loss-efficiency product as small as $aU_\pi L = 1.0$ VdB. The device performance is demonstrated in a data transmission experiment, where we generate on-off-keying (OOK) and four-level pulse-amplitude modulation (PAM4) signals at symbol rates of 100 GBd, resulting in line rates of up to 200 Gbit/s. Bit error ratios are below the threshold for hard-decision forward error correction (HD-FEC) with 7 % coding overhead, leading to net data rates of 187 Gbit/s. This is the highest PAM4 data rate ever achieved for a sub-1 mm silicon photonic MZM.

## 1. Introduction

Electro-optic (EO) Mach-Zehnder modulators (MZM) are key building blocks of optical communication systems. Ideal devices should combine small π-voltages $U_\pi$ with small device lengths $L$ while offering low optical loss in the underlying phase shifters. These quantities are subject to various trade-offs, which can be described by two figures of merit: The π-voltage-length-product $U_\pi L$ and the loss-efficiency product $aU_\pi L$, where $a$ is the phase-shifter propagation loss measured in dB/mm [1,2]. In practical devices, it is challenging to simultaneously minimize both quantities. For example, ultra-low loss-efficiency products of 0.5 VdB were recently achieved in thin-film lithium-niobate (LiNbO$_3$) MZM [3]. However, while these devices offer bandwidths up to 100 GHz and lend themselves to high-speed signaling [3], the efficiency is fundamentally limited by the comparatively low EO coefficient of LiNbO$_3$. This results in rather high $U_\pi L$ products of more than 20 Vmm such that low drive voltages can only be achieved in cm-long devices. Much shorter device lengths can be realized by using semiconductor-based MZM. For example, 4 mm-long indium-phosphide-(InP-)based MZM with bandwidths of 80 GHz were demonstrated [4]. The devices have low $aU_\pi L$ products down to 0.9 VdB, but the $U_\pi L$ products still amount to 6 Vmm, making sub-millimeter InP modulators with low drive voltages hard to realize. In addition, InP-based photonic integrated circuits (PIC) rely on rather expensive fabrication processes on small wafers. Silicon photonic (SiP) devices can overcome this deficiency, exploiting sophisticated high-yield fabrication processes on large-area substrates. However, silicon does not exhibit any Pockels-type second-order nonlinearities due to its inversion-symmetric crystal lattice. Thus, SiP MZM have to rely on phase shifters that exploit the plasma dispersion effect, e.g., by means of reverse-biased pn-junctions that are integrated into the optical waveguides. To increase the efficiency of these phase shifters, high doping concentrations are needed, which increases the optical loss [5]. This trade-off leads to rather high loss-efficiency products, which amount to, e.g., $aU_\pi L$ = 5.8 VdB for best-in-class depletion-type phase shifters, which still feature substantial $U_\pi L$ products of 4.6 Vmm [6]. Much more efficient modulation is achieved in plasmonic-organic hybrid (POH) MZM, which combine plasmonic slot waveguides with highly efficient organic EO materials, and which offer ultra-small $U_\pi L$ products down to 0.05 Vmm [7] with unprecedented bandwidths of hundreds of GHz [8,9]. However, the plasmonic phase-shifter structure is intrinsically linked to strong optical absorption loss, which leads to $aU_\pi L$ products of more than 20 VdB [7]. A modulator technology which simultaneously minimizes both the π-voltage-length product and the loss efficiency product is hence still lacking.

In this paper, we expand on our recent research [10] and show silicon-organic hybrid (SOH) MZM that combine low $U_\pi L$ products of 0.41 Vmm with $aU_\pi L$ products of 1.0 VdB. The MZM rely on 280 µm-long phase shifters and thus offer a small footprint, while the optical insertion loss of the phase shifters amounts to only 0.7 dB. To the best of our knowledge, this is the lowest phase-shifter loss reported so far for a high-speed MZM on the SiP platform. The high-speed performance of the modulator is demonstrated by generating OOK and PAM4 signals at symbol rates of 100 GBd, resulting in a line rate (net data rate) of up to 200 Gbit/s (187 Gbit/s) with a bit error ratio (BER) below the 7% HD-FEC limit. To the best of our knowledge, this is the highest PAM4 data rate so far achieved with a sub-1 mm SiP modulator [11,12]. The low insertion loss and the ability to co-integrate SOH phase shifters with the full portfolio of standard silicon photonic devices makes the concept not only attractive for conventional optical communications, but also for emerging applications in the fields of quantum optics [13,14] or solid-state LiDAR [15].

## 2.   SOH modulator principle

The waveguide structure of SOH modulators is fully compatible with standard fabrication processes of commercial silicon photonic foundries. Specifically, the modulators used in this work were fabricated by UV lithography on 200 mm silicon-on-insulator wafers at A*Star Institute of Microelectronics (IME) in Singapore. Figure 1(a) shows a false-colored top-view micrograph of an SOH MZM. The aluminum (Al) contact pads (yellow) in the top metal layer are connected by vias to a coplanar transmission line in ground-signal-ground (GSG) configuration. The optical path is highlighted in blue: We use grating couplers (GC) to the left and the right of the device to couple light to and from the silicon photonic chip. Strip waveguides connect the GC with the SOH MZM where 2×2 multimode interference couplers (MMI, not visible) equally split and combine the incoming and outgoing light to and from the two arms of the MZM. The MZM arms contain phase shifter sections, which are realized as slot waveguides that are clad with an organic EO (OEO) material (green) in a back-end-of-line post-processing step. Strip-to-slot and slot-to-strip waveguide mode converters (not visible in Fig. 1(a)) are used for an efficient transition between the standard silicon strip waveguides and the slot waveguide at the input and the output of the phase shifter sections [16].

Figure 1(b) shows a schematic cross section of the phase shifter section. The position of the cross section is indicated by the black dashed-dotted line labelled A−A' in Fig. 1(a). The optical slot waveguide is formed by two Si rails with width $w_{rail}$ ≈ 240 nm and height $h_{rail}$ ≈ 220 nm separated by the slot with a width of $w_{slot}$ ≈ 130 nm. The rails are electrically connected to the GSG transmission line in the bottom metal layer by $n$-doped Si slabs with height $h_{slab}$ ≈ 70 nm. A 2 µm-thick buried silicon-dioxide layer (BOX) separates the waveguides from the silicon substrate. The 3 µm thick $SiO_2$ top cladding covers the whole chip and is selectively opened above the phase shifter sections for deposition of the highly efficient OEO material JRD1 [17] using a micro-dispensing tool. The contact pads remain uncovered to ensure reliable contacting with microwave probes. Both the electrical and the optical mode are tightly confined to the OEO material in the slot region, resulting in a strong field overlap and highly efficient phase modulation. The optical confinement to the slot results from the field enhancement in the low-index OEO material at the interface to the high-index Si rails. The electrical mode is confined to the slot, because a signal voltage $U_d$ applied to the GSG transmission line drops entirely across the slot, which is filled by the non-conductive OEO material. This results in large electric driving fields of about $U_d/w_{slot}$ within the slots.

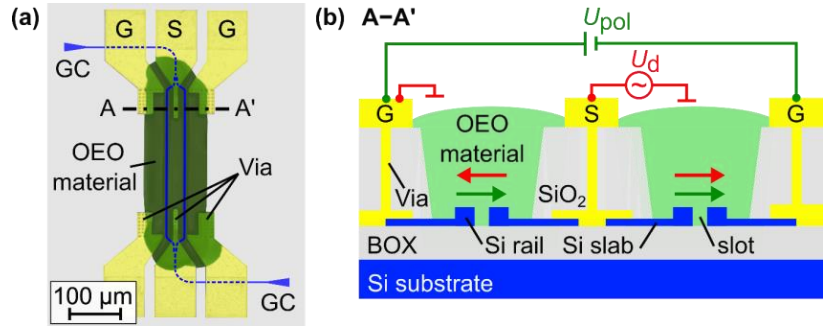

Fig. 1. Silicon-organic hybrid (SOH) modulator concept and structure. **(a)** False-color top-view micrograph of an SOH Mach-Zehnder modulator (MZM). Grating couplers (GC) and optical waveguides to the MZM are drawn in blue. To split and recombine the light in the two arms of the MZM we use multi-mode interference (MMI) couplers, which are hidden below the contact pads (yellow). The electrical signal is applied via a coplanar transmission line in ground-signal-ground (GSG) configuration, from which only the top-layer contact pads are visible. The OEO material (green) is deposited by a micro-dispensing tool and serves as the cladding for the SOH slot waveguide. The black dashed-dotted line labeled A-A' indicates the position of the cross section shown in (b). **(b)** Cross section A-A'. In each arm, two Si rails (width $w_{rail} \approx 240$ nm, height $h_{rail} \approx 220$ nm) define an optical slot waveguide (slot width $w_{slot} \approx 130$ nm), which is filled with the OEO material JRD1 [17]. The rails are electrically connected to the GSG transmission line by doped Si slabs (slab height $h_{slab} \approx 70$ nm). Aluminum vias connect the bottom-layer transmission line to the top-layer contact pads. For poling, the OEO material is heated and a voltage $U_{pol}$ is applied across the floating ground electrodes, thereby aligning the dipolar molecules (green arrows). By cooling the chip, the EO chromophores are frozen in the aligned orientation, and the poling voltage can be removed. After poling, a modulating drive voltage $U_d$ induces electric fields in the slots (red arrows) which are parallel to the poling direction in one arm and anti-parallel in the other arm, thus resulting in efficient push-pull operation of the MZM.

After deposition, the OEO chromophores are randomly oriented, and no macroscopic EO activity can be observed. An average acentric orientation of the molecules and thus a macroscopic EO activity can be induced in a one-time poling process. To this end, we heat the material to the glass transition temperature to increase the molecular mobility. We then apply a DC poling voltage $U_{pol}$ across the floating ground electrodes, which drops across the slots and aligns the dipolar chromophores (green arrows). This results in an average acentric orientation, which is maintained when cooling down the material while keeping the poling field applied. At room temperature the chromophores have lost their mobility, and the poling voltage can be removed. When an RF drive voltage $U_d$ is applied to the GSG transmission line, the electric fields in both arms point in opposite directions (red arrows). This results in a push-pull operation of the MZM, which reduces undesired phase modulation (chirp) of the modulated optical signal [18,19]. Note that the employed OEO material JRD1 is primarily optimized for high efficiency and has a relatively low glass transition temperature of 82 °C [20], which prevents operation at elevated temperatures. However, the chromophore core may be modified by crosslinking agents, which enables lattice hardening after the poling step [21]. For a similar class of OEO material with large EO coefficient of 300 pm/V, this approach resulted in a high glass transition temperature of 160 °C [22] such that thermally induced relaxation of the poling-induced acentric orientation can be neglected at industrially relevant temperatures of 85 °C [22]. An additional challenge is photochemical degradation if the material is exposed to high optical intensities. However, the effect is related to the presence of oxygen such that the OEO material may be protected from degradation by hermetically sealing the modulator by an oxygen blocking sealant [23,24].

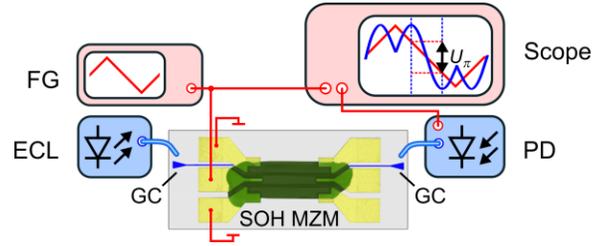

Fig. 2. Setup for measuring the π-voltage $U_\pi$ of the SOH MZM. A low-speed triangular waveform from a function generator (FG) is used to drive the modulator in push-pull mode. Light from an external-cavity laser (ECL) is coupled to the SOH MZM. A photo diode (PD) detects the modulated light. An oscilloscope (Scope) monitors both the drive voltage and the photocurrent. The recorded traces allow to directly read the π-voltage $U_\pi$ [10].

### 3. Determination of π-voltage

To determine the π-voltage of the devices, we use the setup shown in Fig. 2. The SOH MZM is fed by an optical carrier at 1550 nm from an external-cavity laser (ECL), which is coupled to the device via grating couplers (GC). The device is driven by a low-frequency triangular waveform that is provided by a function generator (FG) and coupled to the chip via DC probes. The modulated light is coupled to a fiber and detected by a photodiode (PD), which is connected to an oscilloscope (Scope) to simultaneously monitor the MZM output power and the drive voltage. If the peak-to-peak amplitude of the drive voltage exceeds the π-voltage $U_\pi$, the latter can be directly measured from the drive voltage difference between the minimum and the maximum power transmission of the MZM. The lowest π-voltage measured for the 280 µm-long SOH MZM investigated in this work amounts to 1.48 V, resulting in a $U_\pi L$ product of 0.41 Vmm for a slot width of 130 nm. In a previous report [10] using also the OEO material JRD1, we achieved a slightly better value of 0.32 Vmm for larger slot widths of 150 nm and 190 nm. The slightly reduced modulation efficiency is attributed to the fact that the EO chromophores near the slot walls are mostly oriented parallel to the sidewalls due to surface interactions. As a consequence, the volume fraction of perpendicularly aligned chromophores becomes smaller the narrower the slot is [7].

### 4. Determination of insertion loss of SOH phase shifters

To determine the phase shifter insertion loss, the wavelength of the ECL is swept and the wavelength-dependent total transmitted optical power $P_{tot}(\lambda)$ at the modulator output is measured. In the following, the fiber-to-fiber attenuation of the device is specified in dB and can be calculated by $a_{tot}(\lambda) = -10\log(P_{tot}(\lambda)/P_0)$, where $P_0$ corresponds to the optical launch power. To de-embed the attenuation of the phase shifter sections, we subtract the dB-values of the attenuation of the two grating couplers ($2a_{GC}(\lambda)$), of the feeding strip waveguides ($a_{Strip}$), of the two MMI couplers ($2a_{MMI}(\lambda)$), and of the strip-to-slot and slot-to-strip mode converters ($2a_{Conv}$) from the overall attenuation ($a_{tot}(\lambda)$) of the device. The attenuation of the strip waveguides is calculated according to the specification of the foundry. For all other components, we use suitable test structures on the fabricated wafer, see the Appendix for more details.

The on-chip MZM attenuation $a_{MZM}(\lambda) = a_{tot}(\lambda) - 2a_{GC}(\lambda) - a_{Strip}$ of the bare MZM shows a strong wavelength dependence, see Fig. 3(a). This is caused by the fact that the passive waveguide sections of the MZM arms were designed to have different lengths such that the operating point of the device can be set by the laser wavelength. This leads to a transmission characteristic in which constructive and destructive interference alternates. We choose wavelengths $\lambda_i$ of constructive interference to determine the phase shifter insertion loss $a_{\mathrm{PS}}(\lambda_i)$ by

$$a_{\mathrm{PS}}(\lambda_i) = a_{\mathrm{MZM}}(\lambda_i) - 2a_{\mathrm{MMI}}(\lambda_i) - 2a_{\mathrm{Conv}}. \quad (1)$$

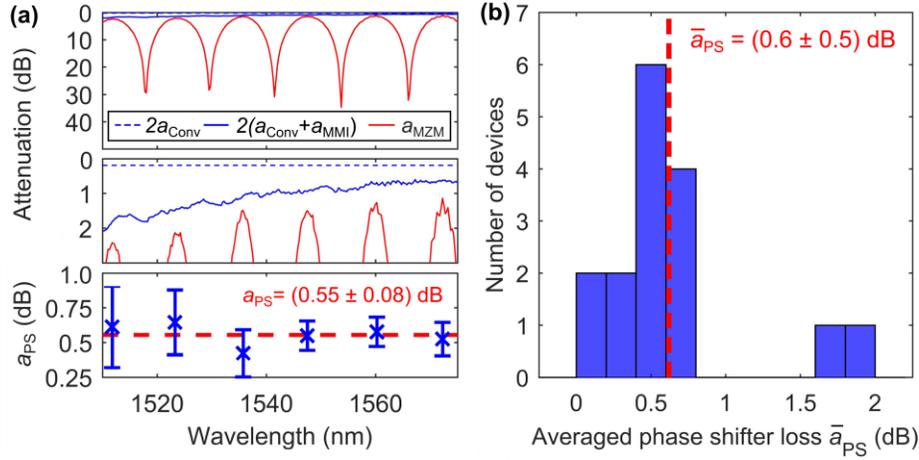

Fig. 3. Measurement of phase-shifter insertion loss. **(a)** Top panel: On-chip attenuation $a_{\mathrm{MZM}}(\lambda)$ of an imbalanced 280 µm-long SOH MZM (red line) as a function of wavelength. The blue solid line represents the attenuation $2a_{\mathrm{MMI}}(\lambda) + 2a_{\mathrm{Conv}}$ of the two MMI couplers and of the two strip-to-slot mode converters. Center panel: Zoom-in of the low-attenuation region seen in the top panel. The blue dashed line indicates the attenuation $2a_{\mathrm{Conv}}$. The envelope of $a_{\mathrm{MZM}}(\lambda)$ follows $2a_{\mathrm{MMI}}(\lambda) + 2a_{\mathrm{Conv}}$. Bottom panel: Calculated phase shifter loss $a_{\mathrm{PS}}(\lambda_i)$ (blue crosses) for one device for various wavelengths $\lambda_i$ where constructive interference occurs. The red dashed line corresponds to an average over the insertion losses $a_{\mathrm{PS}}(\lambda_i)$ and amounts to $\bar{a}_{\mathrm{PS}} = 0.55\,\mathrm{dB}$ with a standard deviation of $\sigma_{a_{\mathrm{PS}}} = 0.08\,\mathrm{dB}$. **(b)** Histogram of wavelength-averaged phase shifter losses $\bar{a}_{\mathrm{PS}}$ of 16 nominally identical devices from four different locations of the wafer. The histogram mean amounts to $\bar{\bar{a}}_{\mathrm{PS}} = 0.6\,\mathrm{dB}$ (red dashed line) and the standard deviation is $\sigma_{\bar{a}_{\mathrm{PS}}} = 0.5\,\mathrm{dB}$. The two outliers are attributed to the fact that the OEO material is filled into the slots by a manual process, and we expect that an automated dispensing of the OEO material will further improve the uniformity and reduce the losses.

The calculated insertion loss $a_{\mathrm{PS}}(\lambda_i)$ varies slightly in the investigated wavelength range between 1510 nm and 1580 nm. We therefore specify a mean phase shifter insertion loss $\bar{a}_{\mathrm{PS}}$ obtained by averaging over all considered wavelengths $\lambda_i$. The procedure is illustrated in Fig. 3(a), where the top panel shows the MZM attenuation $a_{\mathrm{MZM}}(\lambda) = a_{\mathrm{tot}}(\lambda) - 2a_{\mathrm{GC}}(\lambda) - a_{\mathrm{Strip}}$ of a typical imbalanced 280 µm long SOH MZM (red line). We find an extinction ratio of around 30 dB, which is consistent for all investigated devices. The blue solid line indicates the losses $2a_{\mathrm{MMI}}(\lambda) + 2a_{\mathrm{Conv}}$ of the MMI and the strip-to-slot converters in the MZM as obtained from separate test structures on the same wafer. The center panel shows a zoom-in of the low-attenuation region, in which the blue dashed line additionally indicates the converter attenuation $2a_{\mathrm{Conv}}$. The envelope of $a_{\mathrm{MZM}}(\lambda)$ follows $2a_{\mathrm{MMI}}(\lambda) + 2a_{\mathrm{Conv}}$, which indicates that the phase shifter section itself does not show a strong wavelength dependence. The bottom panel displays the phase shifter insertion loss $a_{\mathrm{PS}}(\lambda_i)$ (blue crosses) according to Eq. (1) for one single device. The error bars reflect the overall measurement uncertainty $\sigma_{\mathrm{meas}}$ in dB arising from the uncertainties of all contributing attenuations, see Appendix for details. The insertion loss of the grating couplers increases towards smaller wavelengths, leading to larger error bars due to smaller absolute power levels obtained in the measurement. The red dashed horizontal line corresponds to the wavelength average $\bar{a}_{PS} = 0.55\,\text{dB}$, and the associated standard deviation amounts to $\sigma_{a_{PS}} = 0.08\,\text{dB}$.

To account for statistical variations, we investigate nominally identical dies from four different positions on the wafer, each die containing four nominally identical SOH MZM with 280 µm-long phase shifters. A histogram of the wavelength-averaged phase shifter losses $\bar{a}_{PS}$ of all 16 devices is shown in Fig. 3(b). By averaging over the 16 devices of the histogram, we estimate a mean phase shifter loss of $\bar{\bar{a}}_{PS} = 0.6\,\text{dB}$ and a standard deviation of $\sigma_{\bar{a}_{PS}} = 0.5\,\text{dB}$. Note that the rather high standard deviation is dominated by two outliers, which we attribute to the fact that the OEO material is filled into the slots by a manual process. We expect that an automated dispensing of the OEO material will improve these figures.

The currently used MZM still suffer from a low doping level in the silicon slabs, which leads to high *RC* time constants and hence limits the bandwidth of the phase shifters. In the subsequent 100 GBd signaling experiments, we emulate an increased doping concentration by applying a gate field between the Si substrate and the silicon device layer. This induces an electron accumulation in the Si device layer and thereby increases the slab conductivity [25], but also increases the total optical insertion loss by 0.8 dB for an applied gate field of 0.1 V/nm. Note, however, that this increase in insertion loss comprises the contributions of both the 280 µm-long phase shifter and the 1.3 mm-long access waveguides. Assuming approximately equal carrier-induced propagation losses in both sections, we estimate an increase of the phase shifter loss of approximately 0.14 dB, leading to a still acceptable phase-shifter insertion loss of around (0.7 ± 0.5) dB. To the best of our knowledge, this is the lowest phase-shifter insertion loss so far demonstrated for a high-speed MZM on the SiP platform. Note also that the gate voltage and the associated loss can be avoided by using optimized doping profiles, which rely on high doping concentrations in the slabs outside the core region of the slot waveguide, and which lead to negligible extra insertion losses of below 0.1 dB [26].

The phase shifter loss of $\bar{a}_{PS} = 0.7\,\text{dB}$ leads to a propagation loss coefficient of $a = \bar{a}_{PS} / L = 2.5\,\text{dB/mm}$. Based on theoretical considerations, we believe that these losses are mainly caused by surface roughness of the slot waveguides and that there is still significant potential for further reduction. In fact, slot waveguides with propagation losses down to 0.2 dB/mm were already shown [27]. Using the determined loss coefficient of $a = 2.5$ dB/mm and the measured $U_\pi L$ product of 0.41 Vmm, we calculate a loss-efficiency product of $aU_\pi L = 1.0$ VdB, which is similar to our previously published results of $aU_\pi L = 1.2$ VdB [10]. However, the devices in our previous work were much larger with 1.5 mm long phase shifters and had larger loss coefficients of 3.9 dB/mm resulting in higher phase shifter insertion losses of 6 dB.

## 5. Data transmission experiment

We demonstrate the high-speed performance of the SOH MZM by generating on-off-keying (OOK) and four-level pulse-amplitude modulation (PAM4) signals at symbol rates of 100 GBd. The experimental setup is shown in Fig. 4(a). An arbitrary-waveform generator (AWG) with a bandwidth of 45 GHz delivers the electrical drive signal, which is boosted by an RF amplifier (RF amp) with a bandwidth of 55 GHz. The drive signal is coupled to the open-ended SOH MZM by a microwave probe. The MZM is optically fed by an external-cavity laser (ECL), which is set to an emission wavelength of 1560 nm and an output power of 11.5 dBm. To compensate for the relatively high insertion losses of the grating couplers at the input and the output of the SOH MZM, we use an erbium-doped fiber amplifier (EDFA), which is followed by a band-pass (BP, bandwidth 2 nm) filter to suppress amplified spontaneous emission (ASE) noise. A variable optical attenuator (VOA) is used to adjust the optical power before coupling the modulated signal to a 70 GHz photodiode (PD). The PD output is recorded by a real-time oscilloscope (RTO) having an analog bandwidth of 63 GHz and a sampling rate of 160 GSa/s. The digitized waveforms are post-processed offline. The digital signal processing (DSP) chain includes resampling, timing recovery, blind adaptive time-domain equalization using the Sato algorithm [28] with a filter length of 58 taps, and finally error counting. The signal is a pseudo random binary sequence of length $2^{11}-1$, which is mapped to OOK or PAM4 symbols and encoded onto pulses with a raised-cosine spectrum (roll-off factor $\beta = 0.1$).

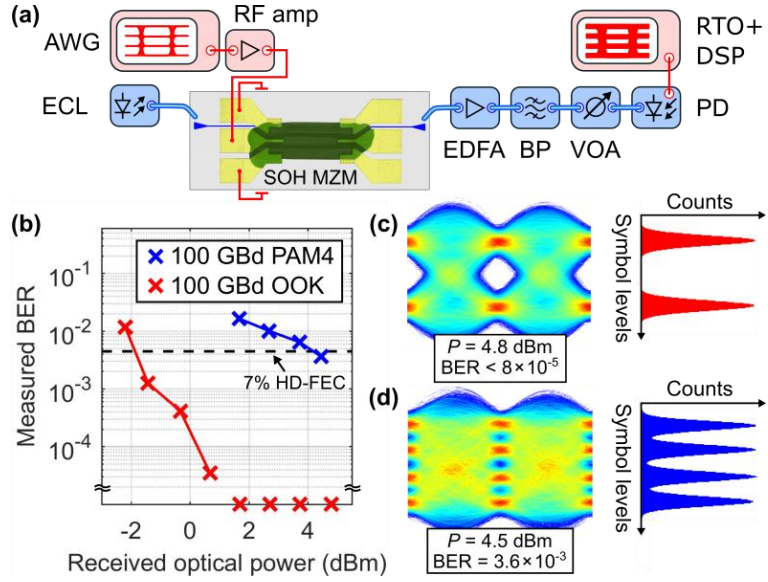

Fig. 4. Data transmission experiments. **(a)** Experimental setup. An external-cavity laser (ECL) provides the optical carrier, which is coupled to the SOH MZM via a grating coupler. An erbium-doped fiber amplifier (EDFA) compensates the optical insertion loss of the grating couplers, and a band pass filter (BP) is used to suppress out-of-band amplified spontaneous emission (ASE) noise. An arbitrary-waveform generator (AWG) is used to drive the open-ended modulator via an RF amplifier (RF amp). A variable optical attenuator (VOA) sets the power level at the receiver, which consists of a high-speed photodiode (PD) connected to a real-time oscilloscope (RTO). Offline digital signal processing (DSP) provides post-equalization and error counting. **(b)** Measured bit error ratio (BER) as a function of the received optical power for OOK and PAM4 at 100 GBd. For OOK (red crosses) and received powers > 1 dBm, we do not detect any errors in our 12.5 µs-long recordings. The BER value remains below the threshold for hard-decision forward error correction (HD-FEC) down to power levels of −1.4 dBm. For PAM4 (blue crosses), we see a power penalty of about 6 dB compared to OOK. At a received power of 4.45 dBm, the BER is $3.6 \cdot 10^{-3}$ and stays just below the threshold for HD-FEC. **(c)**, **(d)** Eye diagrams, corresponding BER, and histograms for 100 GBd OOK and PAM4 signals at the highest measured received power.

For a gate field of 0.1 V/nm the SOH MZM has a 3 dB EO bandwidth of 40 GHz when terminating the electrodes with a 50 Ω resistor. The data transmission experiment, however, is done without device termination. This reduces the bandwidth of the device, but leads to an inherent doubling of the drive voltage through reflection at the open end of the transmission line [29] and thus allows us to use a drive signal with a comparatively small peak-to-peak voltage swing of only 0.72 $V_{pp}$. Such signals can be generated with standard CMOS circuits, thereby enabling highly efficient operation of the devices without any dedicated drive amplifiers [30]. In our experiment, a linear digital pre-equalization of the drive signals is used to flatten the frequency response of the AWG and the RF amplifier. For a 100 GBd PAM4 signal, the peak-to-peak voltage swing at the amplifier output measured at a 50 Ω impedance amounts to 0.78 $V_{pp}$. This reduces to the above-mentioned 0.72 $V_{pp}$ when taking into account the insertion loss of the microwave probe, which amounts to 0.7 dB at the Nyquist frequency of 50 GHz. For the open-ended device, this results in an effective drive voltage of 1.44 $V_{pp}$, which is close to the modulator's DC π-voltage of 1.50 V. However, due to the modulator's frequency roll-off, the MZM is still operated in the linear regime of its transfer function such that nonlinearities do not play a role. The total fiber-to-fiber attenuation of the MZM amounts to 13.6 dB, measured without a gate voltage, and the gate-induced extra loss amounts to 0.8 dB. The wavelength-averaged phase shifter loss amounts to $\bar{a}_{PS} = 0.74\,\text{dB}$ without gate voltage, which increases by an estimated 0.14 dB once the gate is applied.

Figure 4(b) shows the measured bit-error ratio (BER) as a function of the received power. The dashed horizontal line indicates the BER threshold $4.45 \times 10^{-3}$ [31] for hard-decision forward error correction (HD-FEC) with a 7 % overhead. For OOK and received optical powers > 1 dBm, we cannot detect any errors in our 12.5 µs long recordings, which contain $1.25 \times 10^6$ bit. We thus plot the corresponding data points at the lower edge of the diagram. For a received power of −1.4 dBm, the BER amounts to $1.2 \times 10^{-3}$, which is still well below the threshold for HD-FEC. Figure 4(c) depicts the eye-diagram for 100 GBd OOK at a received optical power of 4.8 dBm along with the corresponding histogram of the signal levels measured at the sampling point in the center of the eye.

PAM4 is less robust against inter-symbol interference (ISI) and noise such that our recordings show systematically higher BER values compared to OOK with a power penalty of approximately 6 dB measured at the BER threshold for 7% HD-FEC. For a received optical power of 4.5 dBm, the BER amounts to $3.6 \times 10^{-3}$, which is just below the 7% HD-FEC threshold. In Fig. 4(d), we depict the eye diagram for a received optical power of 4.5 dBm along with the histogram, which was again obtained at the temporal sampling point in the center of the eye diagram. The eyes are not fully open, resulting in histogram counts between the symbol power levels. The achieved line rate amounts to 200 Gbit/s, corresponding to a net data rate of 187 Gbit/s. To the best of our knowledge, this is the highest PAM4 data rate so far demonstrated using a sub-1 mm SiP modulator.

## 6. Summary

We show SOH MZM with $U_\pi L$ products of 0.41 Vmm and $aU_\pi L$ products of 1.0 VdB. The compact devices have 280 µm-long phase shifters with optical insertion losses of only 0.7 dB, which corresponds to the lowest phase-shifter insertion loss reported so far for a high-speed MZM on the silicon photonic (SiP) platform. We demonstrate high-speed optical signaling by generating OOK and PAM4 signals at symbol rates of 100 GBd, resulting in line rates of up to 200 Gbit/s. The measured BER values are below the 7 % hard-decision FEC limit. We believe that compact low-loss MZM with small $U_\pi L$ products are not only interesting for optical communications but may be also useful for applications requiring dense photonic integration of energy-efficient phase shifters such as in optical phased arrays or in the field of quantum optics. The presented device concept can easily be extended to in-phase/quadrature (IQ) modulators [32,33].

**Appendix: Determination of phase shifter insertion loss**

To determine the phase-shifter insertion loss, we investigated SOH devices on nominally identical MZM dies from four different positions of a 200 mm wafer, which we label Die 1, Die 2, Die 3, and Die 4. Each die contains four imbalanced SOH MZM with 280 µm-long phase shifters. For each device, we extract the dB-values of the phase shifter loss $a_{PS}(\lambda_i)$ at wavelengths $\lambda_i$ of constructive interference by subtracting the attenuation of the two grating couplers ($2a_{GC}(\lambda_i)$), the two MMI splitters ($2a_{MMI}(\lambda_i)$), the strip waveguides ($a_{Strip}$) and the two mode converters ($2a_{Conv}$) from the total measured attenuation ($a_{tot}(\lambda_i)$) of the respective device. Subsequently, an average of $a_{PS}(\lambda_i)$ over the set of wavelengths $\lambda_i$ gives the mean phase shifter loss $\bar{a}_{PS}$, see Section 4. For the grating couplers, the MMI splitters, and the mode converters, the attenuations are obtained from measurements of nominally identical reference structures fabricated in the same production run. The propagation loss of the strip waveguides is obtained from specifications of the silicon photonic foundry. Table 1 summarizes the values and the standard deviations of the loss contributions obtained for the various building blocks. Note that these loss contributions are generally wavelength-dependent, which is taken into account when estimating $a_{PS}(\lambda_i)$. For simplicity, Table 1 only specifies the values for the loss contributions $a_{MMI}$, $a_{GC}$, and $a_{GC,2}$ and the associated standard deviations at a fixed wavelength of 1560 nm. The losses $a_{Strip}$ and $a_{Conv}$ are assumed to be wavelength-independent for the investigated wavelength range. For $a_{tot}$, we picked the wavelength $\lambda_i$ of constructive interference that is closest to 1560 nm and indicate the used value of $\lambda_i$ in Table 1.

The measurement of the grating coupler (GC) test structure is particularly important since the associated attenuation is by far the largest of all components. Due to general fabrication tolerances, there are variations in GC performance across the wafer, and we hence measure $a_{GC}$ for each die separately and use a test structure directly next to the respective SOH MZM. Each MZM die also contains a test structure for the MMI insertion loss $a_{MMI}$ consisting of 4, 8, and 12 concatenated MMI couplers, which are accessed via grating couplers. A fit of the losses measured for these coupler sequences allows extracting the attenuation per coupler. This measurement also provides an additional value for the grating coupler attenuation $a_{GC,2}$. For all MZM dies, $a_{GC,2}$ and $a_{GC}$ agree very well, and the mean relative deviations of the associated dB-values do not exceed 3 % in the investigated wavelength range. This result indicates that there is only little intra-die variation of the GC performance and that the losses $a_{GC}$ obtained from the GC test structure can be safely used as a reference. For measuring the insertion losses of the strip-to-slot converters, we use a dedicated test structure on a fifth die, which is obtained from the same wafer as the four MZM dies. The test structure comprises 2, 4, 8, and 10 concatenated pairs of strip-to-slot and slot-to-strip converters. A fit to the measured data gives the attenuation per converter.

The error bounds for the measured attenuations $a_{MMI}$, $a_{Conv}$, and $a_{GC,2}$, are obtained from the least-squares fits of the associated linear model to the respective measurement data. For $a_{Strip}$, we rely on the uncertainty specified by the manufacturer. The error bounds for $a_{tot}$ and $a_{GC}$ correspond to a statistical error obtained by measuring the very same structure three times, starting the alignment of the fibers from independent positions.

The phase shifter losses $a_{PS}(\lambda_i)$ are associated with a wavelength-dependent measurement error $\sigma_{meas}$, corresponding to the blue error bars in Fig. 3(a). For each wavelength $\lambda_i$, this error is determined as the root of the summed error squares of the various contributions according to Eq. (1) of the main text. Note that, for simplicity, Table 1 specifies $\sigma_{meas}$ only for the wavelength $\lambda_i$ of constructive interference, which is closest to 1560 nm. The wavelength-averaged phase shifter loss $\bar{a}_{PS}$ along with the measured standard deviation $\sigma_{a_{PS}}$ of $a_{PS}(\lambda_i)$ is specified in the last column of Table 1.

Table 1. Loss contributions obtained for the various building blocks. The losses $a_{MMI}$, $a_{GC}$, and $a_{GC,2}$ are specified at the wavelength of 1560 nm. The total loss $a_{tot}$ is specified at the wavelength $\lambda_i$ of constructive interference, which is closest to 1560 nm. The quantity $\sigma_{meas}$ reflects the total measurement uncertainty of individual $a_{PS}(\lambda_i)$ taking into account the uncertainties of the individual loss contributions, and $\sigma_{a_{PS}}$ represents the measured standard deviation of $a_{PS}(\lambda_i)$. The value specified for $\sigma_{meas}$ refers to the respective wavelength $\lambda_i$ of constructive interference closest to 1560 nm. All losses are specified in dB.

**Die 1**

| $a_{MMI}$ | $a_{GC}$ | $a_{GC,2}$ | MZM# | $\lambda_i$ (nm) | $a_{tot}(\lambda_i)$ | $\sigma_{meas}(\lambda_i)$ | $\bar{a}_{PS} \pm \sigma_{a_{PS}}$ |
|---|---|---|---|---|---|---|---|
| 0.27 ± 0.01 | 5.52 ± 0.09 | 5.48 ± 0.06 | 1 | 1557 | 13.08 ± 0.23 | 0.34 | 0.75 ± 0.13 |
| | | | 2 | 1554 | 13.63 ± 0.06 | 0.18 | 0.74 ± 0.30 |
| | | | 3 | 1561 | 12.44 ± 0.24 | 0.30 | 0.48 ± 0.16 |
| | | | 4 | 1560 | 12.59 ± 0.05 | 0.20 | 0.55 ± 0.08 |

**Die 2**

| $a_{MMI}$ | $a_{GC}$ | $a_{GC,2}$ | MZM# | $\lambda_i$ (nm) | $a_{tot}(\lambda_i)$ | $\sigma_{meas}(\lambda_i)$ | $\bar{a}_{PS} \pm \sigma_{a_{PS}}$ |
|---|---|---|---|---|---|---|---|
| 0.28 ± 0.01 | 5.14 ± 0.09 | 5.01 ± 0.04 | 1 | 1563 | 12.96 ± 0.06 | 0.12 | 1.78 ± 0.12 |
| | | | 2 | 1561 | 12.81 ± 0.07 | 0.17 | 1.83 ± 0.25 |
| | | | 3 | 1558 | 11.75 ± 0.08 | 0.18 | 0.41 ± 0.14 |
| | | | 4 | 1562 | 11.67 ± 0.08 | 0.14 | 0.46 ± 0.07 |

**Die 3**

| $a_{MMI}$ | $a_{GC}$ | $a_{GC,2}$ | MZM# | $\lambda_i$ (nm) | $a_{tot}(\lambda_i)$ | $\sigma_{meas}(\lambda_i)$ | $\bar{a}_{PS} \pm \sigma_{a_{PS}}$ |
|---|---|---|---|---|---|---|---|
| 0.27 ± 0.01 | 4.36 ± 0.06 | 4.21 ± 0.05 | 1 | 1563 | 10.03 ± 0.06 | 0.11 | 0.51 ± 0.28 |
| | | | 2 | 1563 | 10.21 ± 0.08 | 0.12 | 0.53 ± 0.19 |
| | | | 3 | 1564 | 9.57 ± 0.05 | 0.10 | 0.03 ± 0.31 |
| | | | 4 | 1564 | 9.68 ± 0.27 | 0.28 | 0.21 ± 0.32 |

**Die 4**

| $a_{MMI}$ | $a_{GC}$ | $a_{GC,2}$ | MZM# | $\lambda_i$ (nm) | $a_{tot}(\lambda_i)$ | $\sigma_{meas}(\lambda_i)$ | $\bar{a}_{PS} \pm \sigma_{a_{PS}}$ |
|---|---|---|---|---|---|---|---|
| 0.27 ± 0.01 | 4.93 ± 0.02 | 4.78 ± 0.02 | 1 | 1554 | 12.12 ± 0.07 | 0.10 | 0.64 ± 0.13 |
| | | | 2 | 1555 | 12.23 ± 0.10 | 0.14 | 0.70 ± 0.25 |
| | | | 3 | 1563 | 11.10 ± 0.08 | 0.12 | 0.01 ± 0.36 |
| | | | 4 | 1559 | 11.46 ± 0.07 | 0.11 | 0.30 ± 0.28 |

**Die 5**

| $a_{Conv}$ |
|---|
| 0.10 ± 0.02 |

**Specified by silicon photonic foundry**

| $a_{Strip}$ |
|---|
| 0.32 ± 0.06 |

**Funding**

DFG projects HIPES (383043731), PACE (403188360), GOSPEL (403187440); BMBF project SPIDER (01DR18014A); IARPA SuperCables Program ICENET (W911NF1920114); ERC Consolidator Grant "TeraSHAPE" (773248); Alfried Krupp von Bohlen und Halbach Foundation; Karlsruhe School of Optics and Photonics (KSOP)

**References**

1. X. Tu, T.-Y. Liow, J. Song, M. Yu, and G. Q. Lo, "Fabrication of Low Loss and High Speed Silicon Optical Modulator Using Doping Compensation Method," Opt. Express **19**, 18029 (2011).
2. C. Koos, J. Leuthold, W. Freude, M. Kohl, L. Dalton, W. Bogaerts, A. L. Giesecke, M. Lauermann, A. Melikyan, S. Koeber, S. Wolf, C. Weimann, S. Muehlbrandt, K. Koehnle, J. Pfeifle, W. Hartmann, Y. Kutuvantavida, S. Ummethala, R. Palmer, D. Korn, L. Alloatti, P. C. Schindler, D. L. Elder, T. Wahlbrink, and J. Bolten, "Silicon-Organic Hybrid (SOH) and Plasmonic-Organic Hybrid (POH) Integration," J. Lightw. Technol. **34**, 256–268 (2016).
3. C. Wang, M. Zhang, X. Chen, M. Bertrand, A. Shams-Ansari, S. Chandrasekhar, P. Winzer, and M. Lončar, "Integrated Lithium Niobate Electro-Optic Modulators Operating at CMOS-Compatible Voltages," Nature **562**, 101 (2018).
4. Y. Ogiso, J. Ozaki, Y. Ueda, H. Wakita, M. Nagatani, H. Yamazaki, M. Nakamura, T. Kobayashi, S. Kanazawa, Y. Hashizume, H. Tanobe, N. Nunoya, M. Ida, Y. Miyamoto, and M. Ishikawa, "80-GHz Bandwidth and 1.5-V Vπ InP-Based IQ Modulator," J. Lightw. Technol. 1–1 (2019).
5. J. Witzens, "High-Speed Silicon Photonics Modulators," Proc. IEEE **106**, 2158–2182 (2018).
6. Z. Yong, W. D. Sacher, Y. Huang, J. C. Mikkelsen, Y. Yang, X. Luo, P. Dumais, D. Goodwill, H. Bahrami, G.-Q. Lo, E. Bernier, and J. K. Poon, "Efficient Single-Drive Push-Pull Silicon Mach-Zehnder Modulators with U-Shaped PN Junctions for the O-Band," in *Optical Fiber Communication Conference* (OSA, 2017), p. Tu2H.2.
7. W. Heni, C. Haffner, D. L. Elder, A. F. Tillack, Y. Fedoryshyn, R. Cottier, Y. Salamin, C. Hoessbacher, U. Koch, B. Cheng, B. Robinson, L. R. Dalton, and J. Leuthold, "Nonlinearities of Organic Electro-Optic Materials in Nanoscale Slots and Implications for the Optimum Modulator Design," Opt. Express **25**, 2627 (2017).
8. S. Ummethala, T. Harter, K. Koehnle, Z. Li, S. Muehlbrandt, Y. Kutuvantavida, J. Kemal, P. Marin-Palomo, J. Schaefer, A. Tessmann, S. K. Garlapati, A. Bacher, L. Hahn, M. Walther, T. Zwick, S. Randel, W. Freude, and C. Koos, "THz-to-Optical Conversion in Wireless Communications Using an Ultra-Broadband Plasmonic Modulator," Nat. Photonics **13**, 519–524 (2019).
9. M. Burla, C. Hoessbacher, W. Heni, C. Haffner, Y. Fedoryshyn, D. Werner, T. Watanabe, H. Massler, D. L. Elder, L. R. Dalton, and J. Leuthold, "500 GHz plasmonic Mach-Zehnder modulator enabling sub-THz microwave photonics," APL Photonics **4**, 056106 (2019).
10. C. Kieninger, Y. Kutuvantavida, D. L. Elder, S. Wolf, H. Zwickel, M. Blaicher, J. N. Kemal, M. Lauermann, S. Randel, W. Freude, L. R. Dalton, and C. Koos, "Ultra-High Electro-Optic Activity Demonstrated in a Silicon-Organic Hybrid Modulator," Optica **5**, 739–748 (2018).
11. S. Ummethala, J. N. Kemal, M. Lauermann, A. S. Alam, H. Zwickel, T. Harter, Y. Kutuvantavida, L. Hahn, S. H. Nandam, D. L. Elder, L. R. Dalton, W. Freude, S. Randel, C. Koos, S. Randel, C. Koos, and C. Koos, "Capacitively Coupled Silicon-Organic Hybrid Modulator for 200 Gbit/s PAM-4 Signaling," in *Conference on Lasers and Electro-Optics* (OSA, 2019), p. JTh5B.2.
12. A. Samani, D. Patel, M. Chagnon, E. El-Fiky, R. Li, M. Jacques, N. Abadía, V. Veerasubramanian, and D. V. Plant, "Experimental Parametric Study of 128 Gb/s PAM-4 Transmission System Using a Multi-Electrode Silicon Photonic Mach Zehnder Modulator," Opt. Express **25**, 13252 (2017).
13. J. Wang, S. Paesani, Y. Ding, R. Santagati, P. Skrzypczyk, A. Salavrakos, J. Tura, R. Augusiak, L. Mančinska, D. Bacco, D. Bonneau, J. W. Silverstone, Q. Gong, A. Acín, K. Rottwitt, L. K. Oxenløwe, J. L. O'Brien, A. Laing, and M. G. Thompson, "Multidimensional Quantum Entanglement with Large-Scale Integrated Optics.," Science **360**, 285–291 (2018).
14. X. Qiang, X. Zhou, J. Wang, C. M. Wilkes, T. Loke, S. O'Gara, L. Kling, G. D. Marshall, R. Santagati, T. C. Ralph, J. B. Wang, J. L. O'Brien, M. G. Thompson, and J. C. F. Matthews, "Large-Scale Silicon Quantum Photonics Implementing Arbitrary Two-Qubit Processing," Nat. Photonics **12**, 534–539 (2018).
15. C. V. Poulton, A. Yaacobi, D. B. Cole, M. J. Byrd, M. Raval, D. Vermeulen, and M. R. Watts, "Coherent Solid-State LIDAR with Silicon Photonic Optical Phased Arrays," Opt. Lett. **42**, 4091 (2017).
16. R. Palmer, L. Alloatti, D. Korn, W. Heni, P. C. Schindler, J. Bolten, M. Karl, M. Waldow, T. Wahlbrink, W. Freude, C. Koos, and J. Leuthold, "Low-Loss Silicon Strip-to-Slot Mode Converters," IEEE Photonics J. **5**, 2200409–2200409 (2013).
17. W. Jin, P. V. Johnston, D. L. Elder, A. F. Tillack, B. C. Olbricht, J. Song, P. J. Reid, R. Xu, B. H. Robinson, and L. R. Dalton, "Benzocyclobutene Barrier Layer For Suppressing Conductance in Nonlinear Optical Devices During Electric Field Poling," Appl. Phys. Lett. **104**, 243304 (2014).
18. S. Wolf, H. Zwickel, W. Hartmann, M. Lauermann, Y. Kutuvantavida, C. Kieninger, L. Altenhain, R.

Schmid, J. Luo, A. K.-Y. Jen, S. Randel, W. Freude, and C. Koos, "Silicon-Organic Hybrid (SOH) Mach-Zehnder Modulators for 100 Gbit/s on-off Keying," Sci. Rep. **8**, 2598 (2018).
19. H. Zwickel, S. Wolf, C. Kieninger, Y. Kutuvantavida, M. Lauermann, T. de Keulenaer, A. Vyncke, R. Vaernewyck, J. Luo, A. K.-Y. Jen, W. Freude, J. Bauwelinck, S. Randel, and C. Koos, "Silicon-Organic Hybrid (SOH) Modulators for Intensity-Modulation / Direct-Detection Links With Line Rates of up to 120 Gbit/s," Opt. Express **25**, 23784–23800 (2017).
20. W. Jin, P. V. Johnston, D. L. Elder, K. T. Manner, K. E. Garrett, W. Kaminsky, R. Xu, B. H. Robinson, and L. R. Dalton, "Structure–Function Relationship Exploration for Enhanced Thermal Stability and Electro-Optic Activity in Monolithic Organic NLO Chromophores," J. Mater. Chem. C **4**, 3119–3124 (2016).
21. J. Luo, S. Huang, Z. Shi, B. M. Polishak, X.-H. Zhou, and A. K. Jen, "Tailored Organic Electro-optic Materials and Their Hybrid Systems for Device Applications †," Chem. Mater. **23**, 544 (2011).
22. H. Xu, F. Liu, D. L. Elder, L. E. Johnson, Y. de Coene, K. Clays, B. H. Robinson, and L. R. Dalton, "Ultrahigh Electro-optic Coefficients, High Index of Refraction, and Long-term Stability from Diels-Alder Crosslinkable Binary Molecular Glasses," Chem. Mater. acs.chemmater.9b03725 (2020).
23. R. Dinu, D. Jin, G. Yu, B. Chen, D. Huang, H. Chen, A. Barklund, E. Miller, C. Wei, and J. Vemagiri, "Environmental Stress Testing of Electro–Optic Polymer Modulators," J. Lightw. Technol. **27**, 1527–1532 (2009).
24. Min-Cheol Oh, Hua Zhang, Cheng Zhang, H. Erlig, Yian Chang, B. Tsap, D. Chang, A. Szep, W. H. Steier, H. R. Fetterman, and L. R. Dalton, "Recent Advances in Electrooptic Polymer Modulators Incorporating Highly Nonlinear Chromophore," IEEE J. Sel. Top. Quantum Electron. **7**, 826–835 (2001).
25. L. Alloatti, R. Palmer, S. Diebold, K. P. Pahl, B. Chen, R. Dinu, M. Fournier, J.-M. Fedeli, T. Zwick, W. Freude, C. Koos, and J. Leuthold, "100 GHz Silicon–Organic Hybrid Modulator," Light - Sci. Appl. **3**, e173 (2014).
26. H. Zwickel, S. Singer, C. Kieninger, Y. Kutuvantavida, N. Muradyan, T. Wahlbrink, S. Yokoyama, S. Randel, W. Freude, and C. Koos, "A verified equivalent-circuit model for slotwaveguide modulators," arXiv:2001.02642 [physics.app-ph] (2019).
27. R. Ding, T. Baehr-Jones, W.-J. Kim, B. Boyko, R. Bojko, A. Spott, A. Pomerene, C. Hill, W. Reinhardt, and M. Hochberg, "Low-Loss Asymmetric Strip-Loaded Slot Waveguides in Silicon-on-Insulator," Appl. Phys. Lett. **98**, 233303 (2011).
28. Y. Sato, "A Method of Self-Recovering Equalization for Multilevel Amplitude-Modulation Systems," IEEE Trans. Commun. **23**, 679–682 (1975).
29. S. Koeber, R. Palmer, M. Lauermann, W. Heni, D. L. Elder, D. Korn, M. Woessner, L. Alloatti, S. Koenig, P. C. Schindler, H. Yu, W. Bogaerts, L. R. Dalton, W. Freude, J. Leuthold, and C. Koos, "Femtojoule Electro-Optic Modulation Using a Silicon–Organic Hybrid Device," Light - Sci. Appl. **4**, e255 (2015).
30. S. Wolf, M. Lauermann, P. Schindler, G. Ronniger, K. Geistert, R. Palmer, S. Kober, W. Bogaerts, J. Leuthold, W. Freude, and C. Koos, "DAC-Less Amplifier-Less Generation and Transmission of QAM Signals Using Sub-Volt Silicon-Organic Hybrid Modulators," J. Lightw. Technol. **33**, 1425–1432 (2015).
31. F. Chang, K. Onohara, and T. Mizuochi, "Forward Error Correction for 100 G Transport Networks," IEEE Commun. Mag. **48**, S48–S55 (2010).
32. M. Lauermann, S. Wolf, P. C. Schindler, R. Palmer, S. Koeber, D. Korn, L. Alloatti, T. Wahlbrink, J. Bolten, M. Waldow, M. Koenigsmann, M. Kohler, D. Malsam, D. L. Elder, P. V. Johnston, N. Phillips-Sylvain, P. A. Sullivan, L. R. Dalton, J. Leuthold, W. Freude, and C. Koos, "40 GBd 16QAM Signaling at 160 Gb/s in a Silicon-Organic Hybrid Modulator," J. Lightw. Technol. **33**, 1210–1216 (2015).
33. S. Wolf, H. Zwickel, C. Kieninger, M. Lauermann, W. Hartmann, Y. Kutuvantavida, W. Freude, S. Randel, and C. Koos, "Coherent modulation up to 100 GBd 16QAM using silicon-organic hybrid (SOH) devices," Opt. Express **26**, 220–232 (2018).